\documentclass[12pt]{article}%
\usepackage{amsmath,latexsym}
\usepackage{graphicx}
\usepackage{amsmath}
\usepackage{amsfonts}
\usepackage{amssymb}%
\begin{document}

\title{{Possible existence of time machines in
   a five-dimensional spacetime}}
   \author{
Peter K. F. Kuhfittig*\\  \footnote{kuhfitti@msoe.edu}
 \small Department of Mathematics, Milwaukee School of
Engineering,\\
\small Milwaukee, Wisconsin 53202-3109, USA}

\date{}
 \maketitle

\begin{abstract}\noindent
The idea of constructing a time machine is
not new and even received a boost thanks to
the realization that a traversable wormhole
could be converted to a time machine.  It 
was shown in a recent paper that the 
time-travel paradoxes can be avoided by 
assuming the existence of two or more 
histories or parallel timelines.  Accordingly, 
we start with a physically acceptable model, 
a spacetime that is anti-de Sitter due to an 
extra time-like dimension.  This model allows 
the existence of closed time-like curves, as 
well as two different histories, thereby 
suggesting a possible avoidance of the 
time-travel paradoxes.  By assuming that
the extra dimension is independent of the
radial coordinate, the wormhole retains its
basic geometric properties regardless of its
location. 

\end{abstract}

\section{Introduction}\label{S:introduction}
The idea of constructing a time machine is not new.
One of the first proposals, the van Stockum time
machine \cite{wvS, fT}, starts with an infinitely
long rigidly rotating cylinder of dust surrounded
by a vacuum.  A more recent example is the Gott
time machine \cite{jG}, which employs two spinning
cosmic strings to induce closed time-like curves.
Both are considered unphysical \cite{mV}.  Another
approach is the use of wormholes \cite{MTY},
elaborated on in this paper.  For this approach to
 work, traversable wormholes would have to exist
 and be available when and where needed.

 Wormholes are handles or tunnels connecting
 widely separated regions of our Universe or
 different universes altogether.  These are often
 referred to as intra-universe or inter-universe
 wormholes, respectively.  Since we are primarily
 interested in the former, we will assume that
 an intra-universe wormhole is strictly a
 shortcut for an otherwise ordinary trip through
 spacetime.

 A wormhole may be described by the static and
 spherically symmetric line element \cite{MT}
 \begin{equation}\label{E:line1}
ds^{2}=-e^{2\Phi(r)}dt^{2}+\frac{dr^2}{1-b(r)/r}
+r^{2}(d\theta^{2}+\text{sin}^{2}\theta\,
d\phi^{2}),
\end{equation}
using units in which $c=G=1$.  Here $\Phi=\Phi(r)$
is called the \emph{redshift function}, which
must be everywhere finite to avoid the appearance
of an event horizon.  The function $b=b(r)$ is
called the \emph{shape function}, since it
determines the spatial shape of a wormhole when
viewed, for example, in an embedding diagram
\cite{MT}.  The spherical surface $r=r_0$ is the
\emph{throat} of the wormhole.   The shape
function must satisfy the following conditions:
$b(r_0)=r_0$, $b(r)<r$ for $r>r_0$, and
$b'(r_0)\le 1$, called the \emph{flare-out
condition} in Ref. \cite{MT}.  For a
Morris-Thorne wormhole, this condition can
only be met by violating the null energy
condition (NEC), which states that for the
energy-momentum tensor $T_{\alpha\beta}$,
\begin{equation}
  T_{\alpha\beta}k^{\alpha}k^{\beta}\ge 0
\emph{}\quad\text{for all null
    vectors}\quad k^{\alpha}.
\end{equation}
Matter that violates the NEC is called
``exotic" in Ref. \cite{MT}.  In
particular, for the outgoing null vector
$(1,1,0,0)$, the violation takes on the form
\begin{equation}\label{E:NEC1}
   T_{\alpha\beta}k^{\alpha}k^{\beta}=
   \rho +p_r<0.
\end{equation}
Here $T^t_{\phantom{tt}t}=-\rho$ is the energy
density, $T^r_{\phantom{rr}r}= p_r$ is the
radial pressure, and
$T^\theta_{\phantom{\theta\theta}\theta}=
T^\phi_{\phantom{\phi\phi}\phi}=p_t$ is
the lateral pressure.  A final requirement
is asymptotic flatness: $\text{lim}
_{r\rightarrow\infty}\Phi(r)=0$ and
$\text{lim}_{r\rightarrow\infty}b(r)/r=0$.

A different approach to wormholes can be
found in Kuhfittig \cite{pK}, which assumes
the existence of an extra spatial dimension
leading to the following line element:
\begin{equation}\label{E:line2}
  ds^2=-e^{2\Phi(r)}dt^2 +e^{2\lambda(r)}dr^2+r^2
  (d\theta^2+\text{sin}^2\theta\,d\phi^2)
  +e^{2\mu(r,l)}dl^2.
\end{equation}
Symmetry considerations suggested that the
extra term should have the same exponential
form as the first two terms; $l$ is the extra
coordinate.  We also assume that
$e^{2\lambda(r)}=1-b(r)/r$, as before.

We now turn to a completely different
issue regarding the extra dimension.  If
this dimension is indeed space-like, then
the signature is $-++++$, as in Eq.
(\ref{E:line2}).  It is proposed in Ref.
\cite{WP}, however, that the signature
$--+++$ is in principle allowed, thereby
yielding two time-like components.  In
other words, the line element
\begin{equation}\label{E:line3}
  ds^2=-e^{2\Phi(r)}dt^2-e^{2\mu(r,l)}dl^2
  +e^{2\lambda(r)}dr^2+r^2
  (d\theta^2+\text{sin}^2\theta\,d\phi^2)
\end{equation}
would be consistent with Einstein's theory.
The resulting five-dimensional spacetime
is an example of an anti-de Sitter space
and is characterized by a negative
cosmological constant.  (In fact, in the
absence of matter and energy, the curvature
of a space-like section is negative,
corresponding to hyperbolic non-Euclidean
geometry.)  The most important property of 
an anti-de Sitter space is the existence of 
closed time-like curves due to the existence 
of two timelike components. 

There is little point in considering closed 
timelike curves (or backward time travel) 
without facing the time-travel paradoxes. 
This topic is discussed in more detail in Ref. 
\cite{SW23}.  The only way to avoid these 
paradoxes is by assuming the existence of two 
or more histories or parallel timelines, as 
proposed in Eq. (\ref{E:line3}).  We will 
return to this topic in Sec. \ref{S:closed}.  

\section{Preliminary calculations}
        \label{S:prelim}
We start this section with the five-dimensional 
Einstein field equations in the orthonormal frame:
\begin{equation}
   G_{\hat{\alpha}\hat{\beta}}=
   R_{\hat{\alpha}\hat{\beta}}-\frac{1}{2}R
   g_{\hat{\alpha}\hat{\beta}}
   =8\pi T_{\hat{\alpha}\hat{\beta}};
\end{equation}
so the indeces are 0-4.  We can now write 
\[
   T_{00}=-T^t_{\phantom{tt}t}=\rho\quad
      \text{and}\quad T_{11}=
    T^r_{\phantom{rr}r}= p_r.
\]
Since we wish to follow Ref. \cite{pK}, we
need to rewrite line element (\ref{E:line3})
in the following convenient form:
\begin{equation}\label{E:line4}
  ds^2=-e^{2\Phi(r)}dt^2
  +\frac{dr^2}{1-b(r)/r}+r^2
  (d\theta^2+\text{sin}^2\theta\,d\phi^2)
  -e^{2\mu(r,l)}dl^2.
\end{equation}
To study the effect of the extra time-like
dimension, we choose an orthonormal basis
$\{e_{\hat{\alpha}}\}$ which is dual to the
following 1-form basis:
\begin{equation}\label{E:oneform1}
    \theta^0=e^{\Phi(r)}\, dt,\quad \theta^1
    =[1-b(r)/r]^{-1/2}\,dr,
     \quad\theta^2=r\,d\theta, \quad
      \theta^3=r\,
   \,\text{sin}\,\theta\,d\phi,\quad \theta^4\
    =e^{\mu(r,l)}dl;
\end{equation}
this basis does not depend on the signature.

The connection 1-forms, the curvature 2-forms,
and the components of the Riemann curvature
tensor are derived in Ref. \cite{pK}.  In this
paper, we need only the components of the
Ricci tensor \cite{pK}, listed next.
\begin{multline}
\label{E:Ricci1}  R_{00}=-\frac{1}{2}\frac{d\Phi(r)}{dr}\frac{rb'-b}{r^2}
  +\frac{d^2\Phi(r)}{dr^2}\left(1-\frac{b}{r}\right)\\+
  \left[\frac{d\Phi(r)}{dr}\right]^2\left(1-\frac{b}{r}\right)
  +\frac{2}{r}\frac{d\Phi(r)}{dr}\left(1-\frac{b}{r}\right)\\
  +\frac{d\Phi(r)}{dr}\frac{\partial\mu(r,l)}{\partial r}
  \left(1-\frac{b}{r}\right),
\end{multline}
\begin{multline}\label{E:Ricci2}
   R_{11}=\frac{1}{2}\frac{d\Phi(r)}{dr}\frac{rb'-b}{r^2}
  -\frac{d^2\Phi(r)}{dr^2}\left(1-\frac{b}{r}\right)\\
  -\left[\frac{d\Phi(r)}{dr}\right]^2\left(1-\frac{b}{r}\right)
  +\frac{rb'-b}{r^3}-\frac{\partial^2\mu(r,l)}{\partial r^2}
  \left(1-\frac{b}{r}\right)\\
  +\frac{1}{2}\frac{\partial\mu(r,l)}{\partial r}
  \frac{rb'-b}{r^2}-\left[\frac{\partial\mu(r,l)}{\partial r}
  \right]^2\left(1-\frac{b}{r}\right),
\end{multline}
\begin{equation}\label{E:Ricci3}
   R_{22}=R_{33}=-\frac{1}{r}\frac{d\Phi(r)}{dr}
   \left(1-\frac{b}{r}\right)+\frac{1}{2}\frac{rb'-b}{r^3}
   +\frac{b}{r^3}
   -\frac{1}{r}\frac{\partial\mu(r,l)}{\partial r}
   \left(1-\frac{b}{r}\right),
\end{equation}
\begin{multline}\label{E:Ricci4}
  R_{44}=-\frac{d\Phi(r)}{dr}\frac{\partial\mu(r,l)}{\partial r}
  \left(1-\frac{b}{r}\right)-\frac{\partial^2\mu(r,l)}
  {\partial r^2}\left(1-\frac{b}{r}\right)\\
  +\frac{1}{2}\frac{\partial\mu(r,l)}{\partial r}\frac{rb'-b}{r^2}
  -\left[\frac{\partial\mu(r,l)}{\partial r}\right]^2
  \left(1-\frac{b}{r}\right)
  -\frac{2}{r}\frac{\partial\mu(r,l)}{\partial r}
  \left(1-\frac{b}{r}\right).
\end{multline}
With Eq. (\ref{E:NEC1}) in mind, we now
obtain
\begin{equation}\label{E:NEC2}
   8\pi (\rho +p_r)=[R_{00}-\frac{1}{2}R(-1)]+[R_{11}-
   \frac{1}{2}R(1)]=R_{00}+R_{11}.
\end{equation}
Since we are primarily interested in the
vicinity of the throat, we assume that
$1-b(r_0)/r_0=0$.  So
\begin{equation}\label{E:NEC3}
  \left.\rho +p_r\right|_{r=r_0}
  =\frac{1}{8\pi}\frac{b'(r_0)-1}{r_0^2}
  \left[1+\frac{r_0}{2}\frac{\partial\mu(r_0,l)}{\partial r}
  \right].
\end{equation}
At this point we need to introduce a natural
requirement: we will assume that the extra
dimension is independent of $r$; thus
$\partial\mu(r,l)/\partial r=0$ and Eq.
(\ref{E:NEC3}) becomes
\begin{equation}
   \left.\rho +p_r\right|_{r=r_0}
  =\frac{1}{8\pi}\frac{b'(r_0)-1}{r_0^2}<0,
\end{equation}
which is the usual flare-out condition for
a Morris-Thorne wormhole, indicating that
the NEC has been violated.  So the location
of the wormhole is not restricted, even in the
presence of closed time-like curves.

\section{Closed time-like curves}\label{S:closed}
As noted in the Introduction, we are primarily
interested in wormholes that provide a shortcut
for an already existing path.  We have also seen
that the extra time-like dimension, resulting in
an anti-de Sitter space, assures the existence of
closed time-like curves somewhere.  Since $u(r,l)$
is assumed to be independent of $r$, it is possible
in principle to place a wormhole anywhere, even
along a closed time-like curve.  The result is a
wormhole that functions as a time machine.

One normally associates time travel with causality
violations, which is easy to illustrate by pretending
that the pockets of a pool table are wormholes that
act as time machines.  Suppose a ball is tossed
toward a pocket with the initial direction and
velocity so chosen that, if the ball is undisturbed,
it exits the second mouth of the wormhole in the
past and knocks its younger self off course, thereby
preventing it from even entering the wormhole.
So the above initial conditions would be
forbidden.  According to Refs. \cite{aB} and
\cite{TC}, however, this conclusion is at odds
with classical physics: what could prevent an 
experimenter from tossing a ball in a certain 
direction at a particular velocity?  This 
observation is in line with the Novikov 
self-consistency conjecture \cite{iN}, which also
emphasises that one cannot change the past, i.e., 
one cannot change events that have already taken 
place.  On the other hand, as noted earlier, it 
is shown in Ref. \cite{SW23} that we can avoid 
the time-travel paradoxes by using multiple 
histories or, equivalently, parallel timelines: 
Alice steps into a time machine in history 1 
but emerges from it in history 2, leaving 
history 1 intact. 

This conclusion shows, from a practical standpoint, 
the advantage of using traversable wormholes.  We 
saw at the end of Sec. \ref{S:prelim} that the 
independence of $\mu(r,l)$ of $r$ allows a wormhole 
to be placed anywhere. 

\section{Conclusion}
In this paper, we study backward time travel by
means of traversable wormholes.  We start by
assuming the existence of an extra dimension.
According to Ref. \cite{pW}, the field equations
in terms of the Ricci tensor are
\[R_{AB}=0, \quad A,B=0,1,2,3,4.
\]
This vacuum solution includes the Einstein field
equations  \emph{containing matter}, sometimes
referred to as the induced-matter theory: what
we perceive as matter can be viewed as the
impingement of the higher-dimensional space
onto ours.  According to Ref. \cite{pW}, the
introduction of the extra dimension has
produced many other key insights and should
therefore be seen as an extremely useful
mathematical model, even though the extra
dimension cannot be directly observed.

As noted in Sec. \ref{S:introduction},
Einstein's theory allows the extra dimension
to be space-like or time-like.  If time-like,
the result is a five-dimensional anti-de
Sitter space.  Accordingly, the line
element for the wormhole spacetime becomes
\begin{equation*}
  ds^2=-e^{2\Phi(r)}dt^2
  +\frac{dr^2}{1-b(r)/r}+r^2
  (d\theta^2+\text{sin}^2\theta\,d\phi^2)
  -e^{2\mu(r,l)}dl^2.
\end{equation*}
Because of the extra temporal dimension, the
spacetime contains closed time-like curves.  The
assumption that $u(r,l)$ is independent of $r$
leads to
\begin{equation*}
   \left.\rho +p_r\right|_{r=r_0}
  =\frac{1}{8\pi}\frac{b'(r_0)-1}{r_0^2}<0,
\end{equation*}
the usual flare-out condition at the throat.  So
the wormhole could be placed anywhere.  This
approach necessitates the existence of
traversable wormholes.

Finally, according to Ref. \cite{SW23}, we 
can avoid the time-travel paradoxes if we have 
at least two histories, as proposed in Eq. 
(\ref{E:line3}).

Summarising, to allow backward time travel, 
we need access to traversable wormholes since
a wormhole can in principle be converted
to a time machine.  A second requirement is 
the existence of two or more histories, as 
discussed in Sec. \ref{S:introduction}.  While 
these results seem promising at first, to 
avoid the paradoxes, the traveler would have 
to be able to pass from an old to a new 
history.  Whether this makes backward time 
travel any less problematical is an open 
question.

\end{document}